\documentclass[twocolumn,showpacs,preprintnumbers,amsmath,amssymb,epsfig]{revtex4}

\usepackage{graphicx}
\usepackage{dcolumn}
\usepackage{bm}
\usepackage{epsfig}

\usepackage{float}
\usepackage{amsmath}
\usepackage{epsfig,floatflt}
\usepackage{subfigure}
\usepackage[usenames]{color}

\begin{document}


\title{An improved cosmological model fitting of Planck data
       with a dark energy spike}

\author{Chan-Gyung Park}
\affiliation{
         Division of Science Education and Institute of Fusion Science,  \\
         Chonbuk National University,
         Jeonju 561-756, Republic of Korea }
\date{\today}


\begin{abstract}
The $\Lambda$ cold dark matter ($\Lambda\textrm{CDM}$) model is
currently known as the simplest cosmology model that best describes
observations with minimal number of parameters.
Here we introduce a cosmology model that is preferred over the conventional
$\Lambda\textrm{CDM}$ one by constructing dark energy as the sum of the 
cosmological constant $\Lambda$ and the additional fluid that is designed
to have an extremely short transient spike in energy density during
the radiation-matter equality era and the early scaling behavior with
radiation and matter densities. The density parameter of the additional
fluid is defined as a Gaussian function plus a constant in
logarithmic scale-factor space.
Searching for the best-fit cosmological parameters in the presence of
such a dark energy spike gives a far smaller chi-square value by about
five times the number of additional parameters introduced and narrower
constraints on matter density and Hubble constant compared with the
best-fit $\Lambda\textrm{CDM}$ model.
The significant improvement in reducing chi-square mainly comes from
the better fitting of Planck temperature power spectrum around the
third ($\ell \approx 800$) and sixth ($\ell \approx 1800$) acoustic peaks.
The likelihood ratio test and the Akaike information criterion suggest
that the model of dark energy spike is strongly favored by the current
cosmological observations over the conventional $\Lambda\textrm{CDM}$ model.
However, based on the Bayesian information criterion which penalizes models
with more parameters, the strong evidence supporting the presence of dark
energy spike disappears.
Our result emphasizes that the alternative cosmological parameter estimation
with even better fitting of the same observational data is allowed in the
Einstein's gravity.
\end{abstract}

\maketitle

\section{Introduction}
\label{sec:intro}

It is one of the primary aims in cosmology to find the simplest model
that best describes the astronomical observations. Until now, the best
concordance model with minimal number of parameters is currently known
as the $\Lambda$ cold dark matter ($\Lambda\textrm{CDM}$) model with
the cosmological constant $\Lambda$ as dark energy and CDM as dominant
dark matter \cite{Hinshaw-etal-2012,Planck-2013-Cosmol,Planck-2015-Cosmol}.

Many efforts have been made to develop a cosmology model that is better
than the conventional $\Lambda\textrm{CDM}$ model. Determining
which model is preferred over the other is a problem of model selection.
The standard approach used in the model selection is that one constrains
the new model with data using the likelihood method and checks whether
or not it is supported over the previous best model based on the statistical
criteria. In usual cases, the new candidate cosmology model
has more free parameters than $\Lambda\textrm{CDM}$ model while it often
gives a better fitting of observational data. However, simply adding more
parameters and getting smaller chi-square (or larger likelihood) does
not make the relevant model rank as a better model. In order for a new
cosmology model to be ranked as a model better than $\Lambda\textrm{CDM}$
model, it should pass through at least one of the model selection criteria
such as the likelihood ratio test \cite{LRT}, Akaike information criterion
\cite{Akaike-1974}, Bayesian information criterion \cite{Schwarz-1978},
Bayesian evidence, and so on (see also \cite{Liddle-2004,Saini-etal-2004,
Szydlowski-etal-2015,Tan-Biswas-2012,Heavens-etal-2007,Efstathiou-2008}
for applications in cosmology with brief reviews).

Although many dark energy models have been proposed, most of them give
only a small improvement in fitting the observational data compared
with the $\Lambda\textrm{CDM}$ model and do not pass through the model
selection criteria with high significance, implying that the
$\Lambda\textrm{CDM}$ model is the final winner in the competition of
model selection
\cite{Saini-etal-2004,Efstathiou-2008,Szydlowski-etal-2015,Liddle-etal-2006,
March-etal-2011,Serra-etal-2007,Biesiada-etal-2011,Borowiec-etal-2007,
DeFelice-etal-2012,delCampo-etal-2011,Gong-etal-2011,Lu-etal-2008}.

Recently, Park et al.\ \cite{Park-etal-2014} investigated the observational
effect of the early episodically dominating dark energy based on the
minimally coupled scalar field with the Albrecht-Skordis potential,
where the dark energy density transiently becomes strong during a short
period of time. They show that the presence of the early epdisodic dark
energy can affect the cosmological parameter estimation significantly
and conclude that the recent Planck data strongly favor the
$\Lambda\textrm{CDM}$ model because only a limited amount of dark energy
with episodic nature is allowed. In this paper, we introduce a fluid version
of early transiently dominating dark energy model with the similar episodic
nature. Our dark energy model is designed to have a transient spike in energy
density during an extremely short period and the early scaling behavior
with radiation and matter density.
We show that our dark energy model gives a significant improved
fit to the recent observational data with different parameter constraints
and thus is preferred over the best-fit $\Lambda\textrm{CDM}$ model
based on some model selection criteria.
Through the example of dark energy spike model, we show that the
alternative parameter estimation with even better fitting of the same
observational data is allowed in the Einstein's gravity.

This paper is organized as follows. Sec.\ \ref{sec:spike} describes
the fluid-based dark energy spike model with a transient variation in
dark energy density and presents numerical calculations of background
evolution of this model.
In Sec.\ \ref{sec:obs}, observational effects of the dark energy spike
are investigated using the recent observational data such as the cosmic
microwave background radiation (CMB) data from the Planck satellite and
the baryonic acoustic oscillation (BAO) data from the large-scale
structure surveys. The cosmological parameters constrained with
observations are compared in the presence of or without the dark energy
spike.
In Sec.\ \ref{sec:model}, we compare our dark energy model with the
conventional $\Lambda\textrm{CDM}$ model based on some statistical
criteria used in model selection.
The discussion and conclusion are presented in Sec.\ \ref{sec:conc}.
Throughout this paper, we set $c \equiv 1$ and $8\pi G \equiv 1$.

\section{A fluid-based dark energy spike model}
\label{sec:spike}

The quintessence-based early episodically dominating dark energy model 
proposed in \cite{Park-etal-2014} is on a solid theoretical footing, but
has its limitations.
First, it is not easy to control the onset, strength, and duration of the
transient dark energy because the behavior of dark energy strongly depends
on potential parameters and initial conditions.
Second, the scalar-field based dark energy model theoretically does not
accommodate the crossing of phantom divide ($w=-1$) in the dark energy
equation of state, and thus it is not allowed to consider a {\it spike},
a transient and abrupt variation of dark energy density, which inevitably
induces $w=-1$ crossing.
Here we introduce a fluid model of early dark energy that allows a dark
energy spike and is easy to handle numerically.

\begin{figure*}
\mbox{\epsfig{file=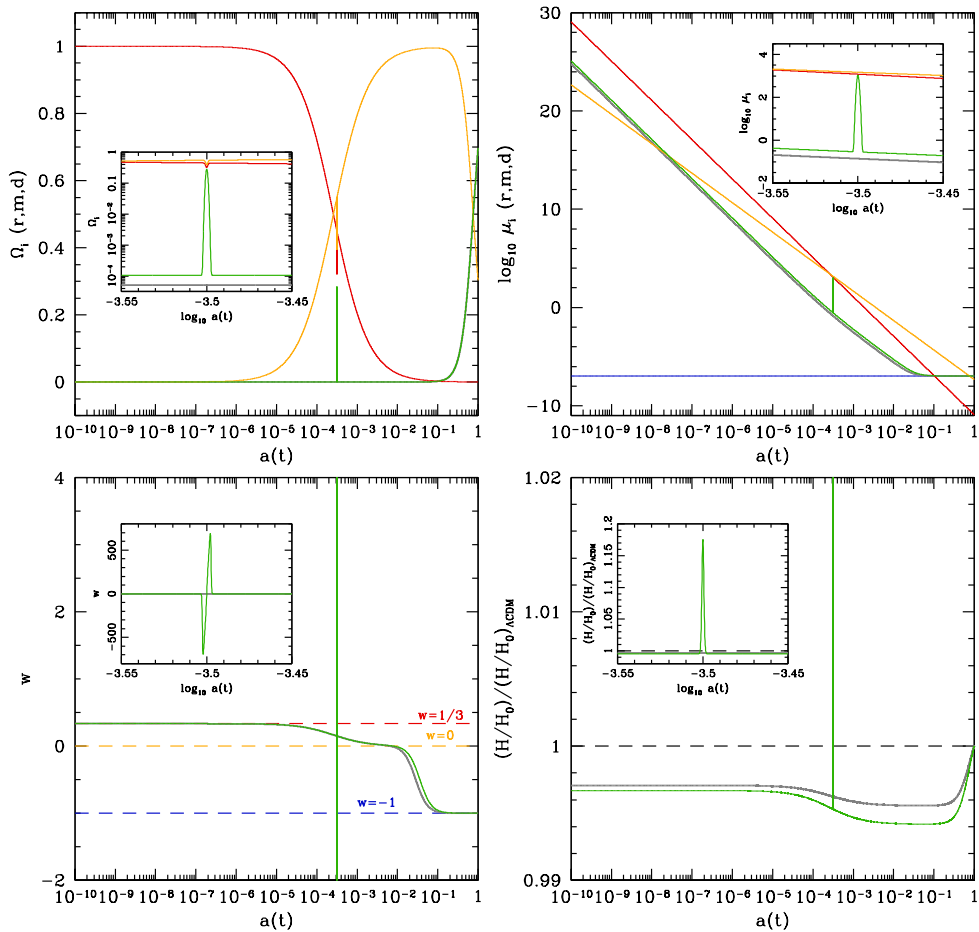,width=130mm,clip=}}
\caption{Evolution of density parameters ($\Omega_i$; $i=r$, $m$, $d$),
         energy densities ($\mu_i$), dark energy equation-of-state ($w$),
         and normalized Hubble parameter divided by the $\Lambda\textrm{CDM}$
         prediction [$(H/H_0)/(H/H_0)_{\Lambda\textrm{CDM}}$]
         in the best-fit spike-DE model with $\log_{10} a_c = -3.5$, $A=0.28$,
         $B=1.1 \times 10^{-4}$, $\sigma=1.5 \times 10^{-3}$.
         In the top-panels, the behaviors of radiation ($r$), matter ($m$),
         and dark energy ($d$) components are shown as red, yellow, and
         green curves, respectively.
         The energy density due to the cosmological constant in the best-fit
         $\Lambda\textrm{CDM}$ model is shown as a blue curve (top-right panel).
         In all panels, grey curves represent the results of scaling-DE model
         with $B=5.2 \times 10^{-5}$.
         The precise values of model parameters used in the numerical
         calculation are presented in Table \ref{tab:best-fit}.
         }
\label{fig:bg_evol}
\end{figure*}

In the conventional $\Lambda\textrm{CDM}$ model, we add a new fluid
(denoted as $x$) with a transient spike in energy density.
We assume that the dark energy density parameter ($\Omega_x$) is represented
as the sum of a Gaussian function and a constant in logarithmic scale-factor
space,
\begin{equation}
   \Omega_x (a)=\frac{\mu_x}{3H^2}
       =A\exp \left[-\frac{(\ln a - \ln a_c)^2}{2\sigma^2} \right] + B,
\label{eq:Omega_X}
\end{equation}
where $\mu_x$ is the energy density of the additional fluid,
$a(t)$ is the cosmic scale factor normalized to unity at present,
$H=\dot{a}/a$ is the Hubble parameter at epoch $a$ with a dot as a
time-derivative, and $A$, $a_c$, and $\sigma$ are interpreted as amplitude
(strength), temporal position, and duration of the dark energy spike,
respectively. The constant term $B$ denotes the level of early dark
energy that exists from the beginning of the universe and is responsible
for the scaling evolution where the dark energy density follows that of
the dominant fluid.
In this paper, we incorporate the $x$-fluid and the cosmological constant
$\Lambda$ into the effective fluid of dark energy (DE).
The behavior of dark energy in our model shows early scaling evolution
with radiation and matter densities with a sudden dark energy spike
at a particular epoch and the late-time acceleration phase due to the
cosmological constant. From here on, we call the dark energy model with
both scaling and transient behaviors as spike-DE model, and the model
only with the scaling behavior as scaling-DE model. 
Note that both the $\Lambda\textrm{CDM}$ model ($A=0$, $B=0$) and
the scaling-DE model ($A=0$ and $B \ge 0$) are nested within the spike-DE model.

In the presence of the $x$-fluid, the squared Hubble parameter normalized
with the present value is given by   
\begin{equation}
   \left(\frac{H}{H_0}\right)^2  
       = \frac{
            \Omega_{r0} a^{-4} + \Omega_{m0} a^{-3} + \Omega_{\Lambda 0}
            + \Omega_{K0} a^{-2}}{1-\Omega_x (a)},
\end{equation}
where the subindices $r$, $m$, $K$ represent the radiation, matter, and
spatial curvature, respectively, and the subindex zero indicates the
present value. We assume that the $x$-fluid satisfies the continuity
equation, $\dot\mu_x=-3H(\mu_x + p_x)$. The pressure of the $x$-fluid
is given by
\begin{equation}
   p_x = 3H^2 \Omega_x \left(-1-\frac{2 \dot H}{3 H^2}
                  - \frac{\Omega_x^\prime}{3 \Omega_x}\right),
\end{equation}
where 
\begin{equation}
     \Omega_x^\prime (a)=\frac{d\Omega_x}{d\ln a}
           = - \frac{\ln a - \ln a_c}{\sigma^2} (\Omega_x-B)
\end{equation}
and 
\begin{equation}
\begin{split}
   \frac{\dot H}{H^2}
     &= \frac{1}{1-\Omega_x} \bigg\{
           \left( -2\Omega_{r0} a^{-4} - \frac{3}{2}\Omega_{m0} a^{-3}
                -\Omega_{K0} a^{-2} \right)  \\
     &~~~~~~~~~~~~~ \times \left(\frac{H_0}{H}\right)^2 
        + \frac{1}{2} \Omega_x^\prime \bigg\}.
\end{split}
\end{equation}
Throughout this paper, we consider the spatially flat universe
($\Omega_{K0}=0$). The equation of state of the effective dark energy
fluid (denoted with a subindex $d$) becomes
\begin{equation}
\begin{split}
   w = \frac{p_d}{\mu_d}
     =-1-\frac{2 \dot{H} \Omega_x}{3H^2(\Omega_x+\Omega_\Lambda)}
            -\frac{\Omega_x^\prime}{3(\Omega_x+\Omega_\Lambda)},
\end{split}
\end{equation}
where $\mu_d=\mu_x + \Lambda$, $p_d=p_x-\Lambda$, and
$\Omega_\Lambda=\Lambda/(3H^2)$.
The dark energy density parameter ($\Omega_d=\Omega_x + \Omega_\Lambda$)
has three asymptotic values:
$\Omega_d \simeq A + B$ at the onset of the dark energy spike ($a = a_c$),
$\Omega_d \simeq B$ before and after the onset ($|a - a_c| \gg 0$) and
before the late-time acceleration ($a \ll a_0$), and $\Omega_{d} \simeq
B+\Omega_{\Lambda}$ during the acceleration phase ($a \approx a_0 \gg a_c$).

Figure \ref{fig:bg_evol} shows the evolution of density parameters,
energy densities, dark energy equation of state, and Hubble parameter
in the spike-DE model where a strong dark energy spike occurs at
$a=10^{-3.5}$ ($\ln a_c =-8.059$), together with those in scaling-DE
and $\Lambda\textrm{CDM}$ models. 
As expected, the dark energy densities of spike-DE (green) and scaling-DE
(grey) models show scaling behaviors following radiation and matter
sequentially. The model parameters have been adopted as the best-fit ones
obtained with the recent observational data (see Sec.\ \ref{sec:obs} for
details). In the spike-DE model, the dark energy equation of state 
experiences a change of about three orders of magnitude with the crossing
of phantom divide twice during the occurrence of dark energy
spike. Considering 95.4\% ($2\sigma$) confidence limits of the Gaussian
shape of the spike, it lasts for about 600 years
($3.153\times 10^{-4} <a < 3.172\times 10^{-4}$).

\section{Observational constraints on the dark energy spike model}
\label{sec:obs}

\begin{table}
\caption{Best-fit cosmological parameters of the spatially flat
         $\Lambda\textrm{CDM}$, scaling-DE, and spike-DE models.
}
\begin{ruledtabular}
\begin{tabular}{lccc}
  Parameter     & $\Lambda\textrm{CDM}$    &  Scaling-DE   &  Spike-DE           \\[1mm]
\hline \\[-2mm]
  $A$                        & $0$   & $0$    & $0.28360$  \\[+1mm]
  $B$                        & $0$   & $5.1736\times 10^{-5}$    & $1.0662\times 10^{-4}$  \\[+1mm]
  $\log_{10} a_c$            &\ldots & \ldots & $-3.5$  \\[+1mm]
  $\sigma$                   &\ldots & \ldots & $1.4604\times 10^{-3}$  \\[+1mm]
\hline \\[-2mm]
  $100\Omega_{b}h^2$         & $2.21632$   & $2.22226$    & $2.19523$  \\[+1mm]
  $\Omega_{c}h^2$            & $0.11827$   & $0.11778$    & $0.11777$  \\[+1mm]
  $h$                        & $0.67929$   & $0.68130$    & $0.68158$  \\[+1mm]
  $\tau$                     & $0.09623$   & $0.08908$    & $0.08873$  \\[+1mm]
  $n_s$                      & $0.96550$   & $0.96487$    & $0.97147$  \\[+1mm]
  $r$                        & $0.00048$   & $0.00046$    & $0.01640$  \\[+1mm]
  $\ln[10^{10}A_s]$          & $3.09985$   & $3.08397$    & $3.08691$  \\[+1mm]
  $t_0$ (Gyr)                & $13.7992$   & $13.7930$    & $13.7983$  \\[+1mm]
\hline \\[-2mm]
  $A^{\textrm{PS}}_{100}$    & $178.3636$  & $138.0731$   & $140.2106$ \\[+1mm]
  $A^{\textrm{PS}}_{143}$    & $62.92783$  & $50.31658$   & $61.88821$ \\[+1mm]
  $A^{\textrm{PS}}_{217}$    & $118.6188$  & $115.4876$   & $126.9649$ \\[+1mm]
  $A^{\textrm{CIB}}_{143}$   & $6.620212$  & $3.852115$   & $5.640900$ \\[+1mm]
  $A^{\textrm{CIB}}_{217}$   & $25.52911$  & $26.94230$   & $23.30611$ \\[+1mm]
  $A^{\textrm{tSZ}}_{143}$   & $3.724382$  & $8.408760$   & $2.995350$ \\[+1mm]
  $r^{\textrm{PS}}_{143\times 217}$
                             & $0.9075909$ & $0.8956619$  & $0.9206412$ \\[+1mm]
  $r^{\textrm{CIB}}_{143\times 217}$
                             & $0.2190109$ & $0.3866272$  & $9.2171337\times 10^{-3}$ \\[+1mm]
  $\gamma^{\textrm{CIB}}$    & $0.5448702$ & $0.5265283$  & $0.5609424$ \\[+1mm]
  $c_{100}$                  & $1.000590$  & $1.000599$   & $1.000575$ \\[+1mm]
  $c_{217}$                  & $0.9963431$ & $0.9962796$  & $0.9968735$ \\[+1mm]
  $\xi^{\textrm{tSZ-CIB}}$   & $0.5315524$ & $5.734012\times 10^{-4}$    & $0.2737207$ \\[+1mm]
  $A^{\textrm{kSZ}}$         & $0.1122116$ & $0.2684244$  & $0.1527581$ \\[+1mm]
  $\beta_1^1$                & $0.5376251$ & $0.5772729$  & $0.2189442$ \\[+1mm]
\end{tabular}
\end{ruledtabular}
\label{tab:best-fit}
\end{table}

We probe the observational signatures of our spike-DE model by considering
both the scalar- and tensor-type perturbations in a system of multiple
components for radiation, matter, and effective dark energy fluid
(the $x$-fluid plus the cosmological constant). For this aim,
we have modified the publicly available CAMB/CosmoMC package
(version of Dec.\ 13 2013) \cite{camb,cosmomc} to include the evolution
of background and perturbation of the effective dark energy fluid,
and explored the allowed ranges of the conventional cosmological
parameters in the presence of a dark energy spike using the Planck CMB
data together with other external data sets.
For the evolution of perturbed density and velocity of the $x$-fluid,
the parametrized post-Friedmann prescription for the dark energy
perturbations is used to allow the multiple crossing of phantom
divide ($w=-1$) in the time-dependent dark energy equation of state
\cite{PPF}. Following the Planck team's analysis, we assume the current
CMB temperature as $T_0=2.7255~\textrm{K}$ and the effective number of
neutrinos as $N_\nu = 3.046$ with a single massive eigenstate of mass
$m_\nu =0.06~\textrm{eV}$ \cite{Planck-2013-Cosmol}.

We use the CMB data obtained with Planck \footnote{An ESA science mission
with instruments and contributions directly funded by ESA Member States,
NASA, and Canada; http://www.esa.int/Planck}, which is a combination of
the CMB temperature anisotropy angular power spectrum up to small angular
scales ($\ell = 2500$) and the Wilkinson Microwave Anisotropy Probe 9-year
polarization data \cite{WMAP9}. We used four Planck CMB likelihood
data sets (version of 2013),
{\tt Lowlike} for low $\ell$ temperature and polarization likelihood
covering $\ell=2$--$32$,
{\tt Commander} for low $\ell$ temperature-only likelihood covering
$\ell=2$--$49$,
{\tt CamSpec} for high $\ell$ temperature-only likelihood with
$\ell=50$--$2500$ \cite{Planck-2013-Like},
and {\tt Lensing} for lensing effect \cite{Planck-2013-Lensing}.
As the external data derived from the large-scale structure observations,
we also have used the BAO data points measured by Sloan Digital Sky Survey
Data Release 7 (DR7) \cite{SDSS-DR7}, Baryon Oscillation Spectroscopic
Survey Data Release 9 (DR9) \cite{BOSS-DR9}, 6dF Galaxy Survey \cite{6dF},
and Wiggle Z surveys \cite{WiggleZ}.

\begin{figure*}
\mbox{\epsfig{file=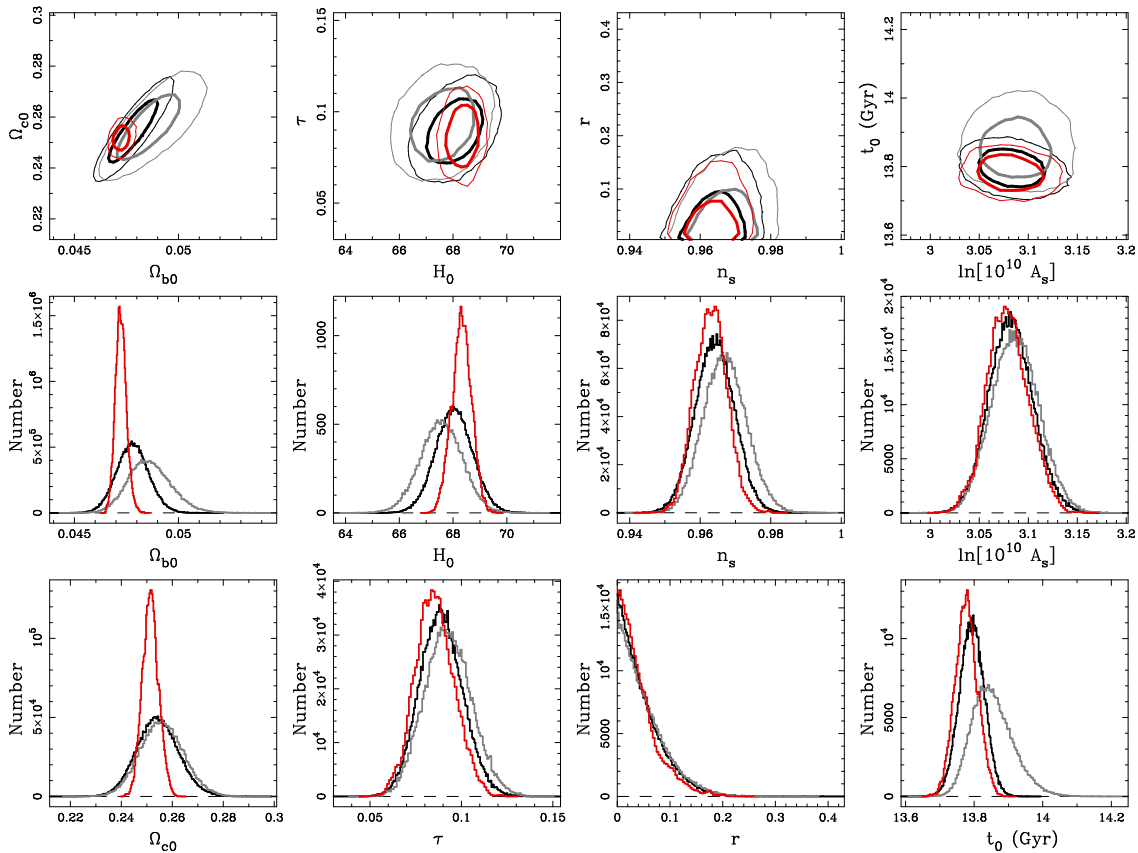,width=150mm,clip=}}
\caption{Top: Two-dimensional likelihood contours of the conventional
         cosmological parameters favored by the Planck CMB and BAO data
         sets for spike-DE (red), scaling-DE (grey), and
         $\Lambda\textrm{CDM}$ (black curves) models.
         For the spike-DE model, dark energy parameters have been fixed
         with the values given in Table \ref{tab:best-fit}.
         The thick and thin solid curves indicate the 68.3\% and 95.4\%
         confidence limits, respectively.
         Middle and bottom: Marginalized one-dimensional likelihood
         distributions for each cosmological parameter, with arbitrary
         normalizations.
         }
\label{fig:like}
\end{figure*}

With CMB and BAO data, we have constrained the parameter space of
the spatially flat $\Lambda\textrm{CDM}$, scaling-DE, and spike-DE models
that are favored by the observations. We limit our investigation by
considering a spike-DE model with a dark energy spike occurring near the
radiation-matter equality era ($\log_{10} a_c = -3.5$);
see Fig.\ \ref{fig:bg_evol}. The reason for choosing such an epoch
is that the transient domination of dark energy near the radiation-matter
equality most affects the evolution of density perturbations in the
scalar-field-based early dark energy model, inducing a highly oscillatory
feature in the angular power spectrum of temperature fluctuations at
high multipoles (see Fig.\ 3 of Ref.\ \cite{Park-etal-2014}).
The free conventional cosmological parameters are
$\Omega_{b0}h^2$, $\Omega_{c0}h^2$, $h$, $\tau$, $n_s$, $r$, and
$\ln[10^{10} A_s]$, where $\Omega_{b0}$ ($\Omega_{c0}$) is the baryon
(CDM) density parameter at the current epoch, $h$ is the normalized
Hubble constant with $H_0=100 h~\textrm{km}~\textrm{s}^{-1}~\textrm{Mpc}^{-1}$,
$\tau$ is the reionization optical depth, $n_s$ is the spectral index
of the primordial scalar-type perturbation, $r$ is the ratio of tensor-
to scalar-type perturbations, and $A_s$ is the amplitude of the primordial
curvature perturbations with $A_s=k^3 P_{\mathcal{R}}(k)/(2\pi^2)$ at
the pivot scale $k_0=0.05~\textrm{Mpc}^{-1}$. The running of spectral
index is not considered.
There are also several foreground and calibration parameters
(see Table \ref{tab:best-fit} and \cite{Planck-2013-Cosmol}
for detailed descriptions of the parameters).

The free parameters of spike-DE model are $A$, $B$, and $\sigma$
with $a_c$ fixed [see Eq.\ (\ref{eq:Omega_X})].
With conventional and dark energy model parameters all freely varying,
the Markov chain Monte Carlo (MCMC) chains are not easily converged
due to multiple local maxima in the multidimensional likelihood distribution.
In this work, instead of obtaining the full converged MCMC chains for
all the free parameters, we search for the best-fit location in the
likelihood distribution by manually running the CosmoMC with an option
{\tt action=2} starting at the local maxima found from the trial MCMC chains
obtained with all parameters varying.
The results are summarized in Table \ref{tab:best-fit}, which lists
the parameters of $\Lambda\textrm{CDM}$, scaling-DE, and spike-DE models
that best describe the observational data, together with the cosmic age
($t_0$) and the parameters related with foregrounds and instrumental
calibrations.

\begin{table}
\caption{Mean and standard deviation ($68.3$\% confidence limit) of
         the conventional cosmological parameters estimated from the
         marginalized one-dimensional likelihood distribution for best-fit
         $\Lambda\textrm{CDM}$, scaling-DE, spike-DE models
         constrained with the Planck CMB and BAO data sets.
         For the tensor-to-scalar ratio $r$ and the level of early dark
         energy $B$, the upper limits are presented.
}
\begin{ruledtabular}
\begin{tabular}{lccc}
  Parameter     & $\Lambda\textrm{CDM}$     &  Scaling-DE          &  Spike-DE            \\[1mm]
\hline \\[-2mm]
  $100\Omega_{b0}$   & $4.784  \pm 0.076$   & $4.870  \pm 0.102$    & $4.727  \pm 0.026$    \\[+1mm]
  $\Omega_{c0}$      & $0.2547 \pm 0.0081$  & $0.2564 \pm 0.0084$   & $0.2518 \pm 0.0031$   \\[+1mm]
  $h$                & $0.6807 \pm 0.0069$  & $0.6761 \pm 0.0077$   & $0.6836 \pm 0.0037$    \\[+1mm]
  $\tau$             & $0.090  \pm 0.012$   & $0.093  \pm 0.013$    & $0.086  \pm 0.011$    \\[+1mm]
  $n_s$              & $0.9644 \pm 0.0056$  & $0.9675 \pm 0.0061$   & $0.9635 \pm 0.0046$    \\[+1mm]
  $r$                & $< 0.054$            & $< 0.061 $            & $< 0.048$            \\[+1mm]
  $\ln[10^{10}A_s]$  & $3.084  \pm 0.022$   & $3.088  \pm 0.024$    & $3.081  \pm 0.021$    \\[+1mm]
  $t_0$ (Gyr)        & $13.795 \pm 0.036$   & $13.858 \pm 0.060$    & $13.781 \pm 0.033$      \\[+1mm]
\hline \\[-2mm]
  $B$                & \ldots               & $<0.0022$              & \ldots      \\[+1mm]
\end{tabular}
\end{ruledtabular}
\label{tab:cl}
\end{table}

To see how the conventional cosmological parameters are affected by
the presence of the dark energy spike, we apply the MCMC method to randomly
explore the parameter space that is favored by observations.
For the spike-DE model, we have fixed dark energy parameters $A$, $B$,
and $\sigma$ with the best-fit values given in Table \ref{tab:best-fit}.
For the scaling-DE model, however, the parameter $B$, the initial level
of early dark energy, has been freely varied.  
Figure \ref{fig:like} shows two-dimensional likelihood contours and 
marginalized one-dimensional likelihood distributions of conventional
cosmological parameters favored by the Planck CMB and BAO data sets 
for the spike-DE, scaling-DE, and $\Lambda\textrm{CDM}$ models, estimated
from the converged MCMC chains.
Note that here we present the likelihood distributions of $\Omega_{b0}$
and $\Omega_{c0}$ instead of $\Omega_{b0} h^2$ and $\Omega_{c0} h^2$.
Table \ref{tab:cl} summarizes the mean and 68.3\% confidence limit of
cosmological parameters estimated from the one-dimensional likelihood
distributions.
For the tensor-to-scalar ratio $r$ and the level of early dark energy
$B$, the upper limits (68.3\%) are given. 
Interestingly, compared with $\Lambda\textrm{CDM}$ model,
the spike-DE model gives narrower parameter constraints
on baryon, CDM density parameters and Hubble constant 
with the standard deviations smaller by a factor of $2.9$, $2.6$, $1.9$,
respectively, and best-fit values slightly deviating from those of
$\Lambda\textrm{CDM}$ model. 
Since the likelihoods of the spike-DE model sufficiently overlap with
the $\Lambda\textrm{CDM}$ ones, the estimated parameters of both models
are still consistent with each other.

For the scaling-DE model, the parameter constraints are consistent
with those of $\Lambda\textrm{CDM}$ model, except for
slightly larger values of baryon density and cosmic age.
In this model we have set $B$ as a free parameter 
to constrain the level of early dark energy density ($\Omega_e = B$).
The allowed range for the early dark energy is $\Omega_e < 0.0045$
($95.4$\% confidence limit), which is narrower than the Planck constraint
on the fluid-based early dark energy density parameter of
Doran \& Robbers \cite{Doran-2006} ($\Omega_e < 0.009$;
\cite{Planck-2013-Cosmol}).
Recently, a substantial improvement on the constraint 
$\Omega_e < 0.0036$ (at 95\% confidence level) has been obtained by
the Planck 2015 data analysis \cite{Planck-2015-DE}.

\section{Model comparison}
\label{sec:model}

In this section, we compare the best-fit $\Lambda\textrm{CDM}$, scaling-DE,
spike-DE models to see which model is preferred over the other by the
current observations based on the statistical criteria such as 
the likelihood ratio test and the Akaike and Bayesian information criteria
that are widely used in the model selection.

\begin{table}
\caption{Chi-square ($\chi^2$) values of best-fit $\Lambda\textrm{CDM}$,
         scaling-DE, and spike-DE models, together with
         differences of chi-square ($\Delta\chi^2$),
         Akaike information criterion ($\Delta\textrm{AIC}$), and
         Bayesian information criterion ($\Delta\textrm{BIC}$) relative
         to $\Lambda\textrm{CDM}$ value,
         and $p$-values estimated from the likelihood ratio test (LRT)
         statistic.
}
\begin{ruledtabular}
\begin{tabular}{lrrr}
  Data  & $\Lambda\textrm{CDM}$     &  Scaling-DE  & Spike-DE \\[1mm]
\hline \\[-2mm]
  {\tt Lowlike}     & $2014.578$     & $2014.178$   & $2014.092$ \\[+1mm]
  {\tt Commander}   & $-7.304$       & $-7.471$     & $-8.096$ \\[+1mm]
  {\tt CamSpec}     & $7795.773$     & $7796.223$   & $7777.669$ \\[+1mm]
  {\tt Lensing}     & $9.892$        & $9.190$      & $9.881$ \\[+1mm]
  DR7               & $0.858$        & $0.620$      & $0.439$ \\[+1mm]
  DR9               & $0.431$        & $0.603$      & $0.812$ \\[+1mm]
  6dF               & $0.019$        & $0.034$      & $0.036$ \\[+1mm]
  Wiggle Z          & $0.047$        & $0.024$      & $0.021$ \\[+1mm]
\hline \\[-2mm]
  Total $\chi^2$
                    & $9814.295$     & $9813.400$   & $9794.753$ \\[+1mm]
\hline \\[-2mm]
  $\Delta\chi^2$
                    & \ldots~~       & $-0.895$     & $-19.542$ \\[+1mm]
 $p$-value (LRT)    & \ldots~~       & $0.3441$     & $6.148\times 10^{-4}$\\[+1mm]
 $\Delta\textrm{AIC}$  & \ldots~~    &  $1.105$     & $-11.542$\\[+1mm]
 $\Delta\textrm{BIC}$  & \ldots~~    &  $6.982$     & $11.968$\\[+1mm]
\end{tabular}
\end{ruledtabular}
\label{tab:chi-square}
\end{table}

Table \ref{tab:chi-square} lists the separate chi-square ($\chi^2$) values for
each likelihood data set used in the parameter estimation
for the best-fit $\Lambda\textrm{CDM}$, scaling-DE, and spike-DE models
(see Table \ref{tab:best-fit} for the best-fit values).
The negative chi-square values for {\tt Commander} likelihood data
appear due to the arbitrary normalization of log-likelihood
in the CosmoMC software \footnote{http://cosmocoffee.info/viewtopic.php?t=2216}.
We note that the best-fit cosmological parameters in the presence of
a dark energy spike near the radiation-matter equality gives a far smaller
chi-square value than those of the $\Lambda\textrm{CDM}$ model
by about five times the number of new free parameters of the spike-DE model
($A$, $B$, $\log_{10} a_c$, $\sigma$) with a difference
$\Delta\chi^2 \equiv \chi^2-\chi^2_{\Lambda\textrm{CDM}}=-19.542$,
which is a significant improvement of data-fitting.

The three dark energy models considered here are nested
in the sense that the $\Lambda\textrm{CDM}$ and scaling-DE models are
special cases of the spike-DE model. In this case, we can apply the likelihood
ratio test (LRT) as a model selection method, where the null model is the
$\Lambda\textrm{CDM}$ model and the alternative model is the scaling-DE
or spike-DE model \cite{LRT,Szydlowski-etal-2015}.
The test statistics is defined as the twice the natural 
logarithm of the ratio of likelihoods of the null and alternative hypotheses
(models) and is equivalent to a difference of chi-square relative to
the $\Lambda\textrm{CDM}$ one,
\begin{equation}
   Q=2 \ln \frac{\mathcal{L}(H_{\Lambda\textrm{CDM}}|D)}{\mathcal{L}(H|D)}
    = \Delta\chi^2,
\end{equation}
where $\mathcal{L}(H|D)$ indicates the maximum likelihood of the alternative
model ($H$) given the data ($D$), and likewise for the null model
($H_{\Lambda\textrm{CDM}}$).
The LRT statistic is a computationally cheap version of the Bayes factor
which provides a criterion for penalizing models with more parameters
based on the Bayesian theory \cite{Saini-etal-2004}.
The test statistic $Q$ can be approximated as the $\chi^2$-distribution
with degrees of freedom ($df$) defined as the additional number
of parameters of the nesting model
($df=4$ for spike-DE model, and $df=1$ for scaling-DE model).
The $p$-value, the probability that the null hypothesis is supported by
the observational data over the alternative one, is calculated from
the cumulative $\chi^2$-distribution and presented in
Table \ref{tab:chi-square}. We find that the $p$-value for the spike-DE
model as alternative is quite small ($p=6.1\times 10^{-4}$),
suggesting a strong preference to the spike-DE model against the
$\Lambda\textrm{CDM}$ model.

The Akaike information criterion (AIC) is defined as \cite{Akaike-1974,
Liddle-2004,Szydlowski-etal-2015}
\begin{equation}
   \textrm{AIC}=-2\ln\mathcal{L} + 2k,
\end{equation}
where $k$ is the number of free parameters of the model considered.
If the alternative model gives a smaller AIC compared with the null 
($\Lambda\textrm{CDM}$) model, it is ranked as a better model
because the discrepancy with the true model is considered to be smaller.
It is generally accepted that the AIC difference of 5 or more gives
a strong evidence supporting the model with smaller AIC value
(see \cite{Tan-Biswas-2012} for the reliability of the AIC method
in cosmological model selection).
The differences of AIC relative to the $\Lambda\textrm{CDM}$ model
($\Delta\textrm{AIC}=\Delta\chi^2 + 2 df$) are listed in Table
\ref{tab:chi-square}.
The scaling-DE model has a positive value of $\Delta\textrm{AIC}=1.1$,
which means that introducing the scaling dark energy without a spike
into the $\Lambda\textrm{CDM}$ model does not improve the fit much.
On the other hand, the negative value of $\Delta\textrm{AIC}=-11.5$
for the spike-DE model
suggests that the alternative model of dark energy with early scaling behavior
and a dark energy spike (near the radiation-matter equality) is strongly
favored by the current cosmological observations over the
$\Lambda\textrm{CDM}$ model.

As an alternative to the AIC, the Bayesian information criterion (BIC) is
often used for model selection, which assigns a conservative penalty 
for large sample size. The BIC is defined as \cite{Schwarz-1978,Liddle-2004,
Szydlowski-etal-2015}
\begin{equation}
   \textrm{BIC}=-2\ln\mathcal{L} + k \ln N,
\end{equation}
where $N$ is the number of data points. We set $N=2637$ 
for Planck+BAO data sets ($31\times 4+48+2451+8+6$ 
for {\tt Lowlike} [TT, TE, EE, BB], {\tt Commander}, {\tt CamSpec},
{\tt lensing}, and BAO data, respectively) to calculate the difference
of BIC relative to the $\Lambda\textrm{CDM}$ value
($\Delta\textrm{BIC}=\Delta\chi^2+ df \ln N$), which are listed in Table
\ref{tab:chi-square}.
Contrary to the AIC result, the spike-DE model has a positive value of
$\Delta\textrm{BIC}=12.0$.
The strong evidence supporting the presence of dark energy spike by the AIC
has disappeared, since the BIC penalizes complex models by the large number of
data points as in the CMB observation.
In the context of BIC, the scaling-DE model with $df=1$ is preferred over
the spike-DE model ($df=4$).

In summary, comparison of the maximum likelihoods of spike-DE and
$\Lambda\textrm{CDM}$ models according to LRT and AIC suggests that
the spike-DE model is strongly preferred over the $\Lambda\textrm{CDM}$ one
while the BIC still indicates the observational data supports the simple
$\Lambda\textrm{CDM}$ model over others.
From the definition of AIC and BIC we see that the AIC is inclined to
select the model that better fits to the data while the BIC selects a simpler
model with less parameters.
Apart from the model selection between $\Lambda\textrm{CDM}$ and spike-DE
models,
at least we conclude that the spike-DE model fits the observational data
far better than $\Lambda\textrm{CDM}$ model with the different cosmological
parameter estimation.

\begin{figure}
\mbox{\epsfig{file=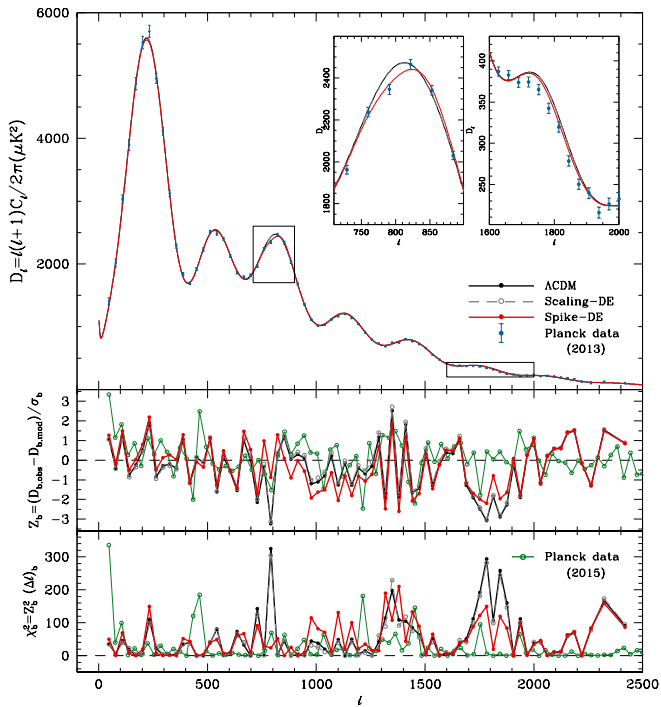,width=87mm,clip=}}
\caption{Top: The CMB temperature angular power spectra of the best-fit
         $\Lambda\textrm{CDM}$ (black), scaling-DE (grey), and
         spike-DE models (red curves).
         The angular power spectrum is given as
         $D_{\ell}=\ell (\ell+1) C_l / (2\pi)$ in $\mu\textrm{K}^2$ unit.
         The Planck 2013 data points are shown with blue dots with error bars.
         Small panels inside magnify regions of $\ell \approx 800$ and $1800$.
         Middle and Bottom: the difference between observation and best-fit
         model prediction of band power divided by the measurement error
         ($\sigma_b$) for each $\ell$-bin (denoted as $b$),
         $Z_b=(D_{b,\textrm{obs}}-D_{b,\textrm{mod}})/\sigma_b$,
         and the sum of contribution due to $Z_b^2$ within the bin,
         $\chi_b^2=Z_b^2 (\Delta\ell)_b$.
         For comparison, the quantities $Z_b$ and $\chi_b^2$ estimated from
         the Planck 2015 data and the best-fit $\Lambda\textrm{CDM}$ model
         prediction (constrained with Plank 2015 TT+LowP+Lesing data)
         are presented (green open circles; \cite{Planck-2015-Cosmol}).
         }
\label{fig:powspec}
\end{figure}

According to Table \ref{tab:chi-square}, the spike-DE model improves fitting
to the {\tt CamSpec} high $\ell$ temperature likelihood data.
Figure \ref{fig:powspec} verifies that the chi-square decrease is
mainly due to the better fitting of Planck temperature power spectrum data
around the third ($\ell\approx 800$) and sixth ($\ell \approx 1800$) acoustic
peaks where strong residuals relative to the best-fit $\Lambda\textrm{CDM}$
model are seen. In the middle and bottom panels of Fig.\ \ref{fig:powspec},
we plot the difference of the observed and the model-predicted band power spectra
($D_b$) normalized with the measurement error ($\sigma_b$) for each $\ell$-bin
(here denoted as $b$),
$Z_b = (D_{b,\textrm{obs}}-D_{b,\textrm{mod}})/\sigma_b$,
and the sum of contribution due to $Z_b^2$ within the bin,
$\chi_b^2 \equiv Z_b^2 (\Delta l)_b$, which approximates the chi-square
contribution for each bin.
The model band power $D_{b,\textrm{mod}}$ is the average of the
CMB angular power spectrum $D_{\ell}=\ell (\ell+1) C_{\ell} / (2\pi)$
predicted by a model within a specified bin.
We use the same $\ell$-bins used in the Planck team's analysis.
It was originally reported that the strong residuals seen around 
the third ($\ell \approx 800$) and fifth ($\ell \approx 1300$--$1500$)
acoustic peaks are real features of the primordial CMB sky
\cite{Planck-2013-Cosmol}.
We note that the best-fit spike-DE model significantly alleviates
the strong residuals around the third ($\ell \approx 800$) and sixth
($\ell \approx 1800$) acoustic peaks observed in the best-fit
$\Lambda\textrm{CDM}$ model (red and black dots).
However, the residual around the fifth peak ($\ell \approx 1300$--$1500$)
still remains in both models.

\section{Discussion and Conclusion}
\label{sec:conc}

In this paper, we investigate the observational effect of the early
episodically dominating dark energy which accommodates
a dark energy spike, a sudden transient variation in dark energy density,
together with early scaling behaviors and late-time acceleration.

The dark energy model with a spike (spike-DE model) near the radiation-matter
equality era improves the fit to the Planck CMB temperature power spectrum
data around the third ($\ell \approx 800$) and sixth ($\ell \approx 1800$)
acoustic peaks.
Comparing the likelihood distributions based on the maximum likelihood ratio
test and the Akaike information criterion as the statistical model selection
methods, we find that the spike-DE model is strongly favored by observations
over the conventional $\Lambda\textrm{CDM}$ model.
Furthermore, the spike-DE model provides the different cosmological parameter
estimation with tighter constraints on matter density and Hubble constant
(see Fig.\ \ref{fig:like} and Table \ref{tab:cl}). 
However, the strong evidence supporting the presence of the dark energy
spike disappears based on the the Bayesian information criterion which
assigns a conservative penalty to the model with a large number of parameters.

We have checked that including high-$\ell$ CMB data observed by the South
Pole Telescope and the Atacama Cosmology Telescope \cite{SPT,ACT} 
or excluding the tensor-type perturbation do not affect our main results.
Besides, we infer that the foreground and instrumental calibration
parameters do not play a major role in improving the fit to the data.
If they do, a significant reduction of chi-square should be seen in the
case of $\Lambda\textrm{CDM}$ model, too.

Very recently, the Planck 2015 data have been publicly available.
The main scientific conclusions of Planck 2015 data analysis are consistent
with the previous results, with cosmological parameters deviating
less than $0.7\sigma$ \cite{Planck-2015-Cosmol}.
As shown in Fig.\ \ref{fig:powspec}, the strong residuals around
$\ell \approx 800$ and $\ell \approx 1800$ in the Planck 2013
temperature power spectrum data are not observed in the Planck 2015
data; the deviation from the best-fit $\Lambda\textrm{CDM}$ model prediction
becomes quite smaller (see green open circles in the middle and bottom panels
of Fig.\ \ref{fig:powspec}).
Therefore, the success of the spike-DE model in improving the fit to
the Planck 2013 temperature power spectrum at those regions is not expected
any more in the recent Planck 2015 data.
However, the presence of strong residuals at $\ell=400$--$500$ and
$\ell \approx 1200$ in the Planck 2015 data still leaves open the possibility
that the new data are fitted by another candidate model of dark energy far
better than $\Lambda\textrm{CDM}$ model.

Through an example of the dark energy spike model, we emphasize that 
the alternative cosmological parameter estimation is allowed in the
Einstein's gravity, with even better fitting of the same observational
data than the conventional $\Lambda\textrm{CDM}$ model.


%
%
\acknowledgements
The author would like to thank Professor Jai-chan Hwang for valuable discussions. 
C.G.P. was supported by Basic Science Research Program through the National
Research Foundation of Korea (NRF) funded by the Ministry of Science, ICT
and Future Planning (No.\ 2013R1A1A1011107).

\def\and{{and }}


\end{document}